\documentclass[12pt,pra,aps,amssymb,amsfonts,amsmath,tightenlines]{revtex4}
\usepackage{dsfont}
\usepackage{graphicx}
\usepackage{amssymb,amsfonts,amsthm}
\usepackage{color}
\usepackage{verbatim}
\usepackage[normalem]{ulem}
\usepackage{enumerate}
\usepackage{mathrsfs}
\usepackage{bm,float}
\usepackage{amsmath}
\DeclareMathOperator*{\argmax}{argmax}

\usepackage{textgreek}
\usepackage{lgreek}

\usepackage{caption}
\newtheorem{Proposition}{Proposition}
\newtheorem{Definition}{Definition}
\usepackage{nicefrac}

\newcommand{\half}{\mbox{$\textstyle \frac{1}{2}$}}

\newcommand{\re}{\mbox{$\rm e$}}

\usepackage{tikz}
\usetikzlibrary{shapes.geometric}
\begin{document}
\title{Valuation of a Financial Claim Contingent on the \\Outcome of a Quantum Measurement}
\author{Lane~P.~Hughston$^1$ and Leandro S\'anchez-Betancourt$^{2,\, 3}$}

\affiliation{
$^1$Department of Computing, Goldsmiths University of London\\ New Cross, London SE14\,6NW, United Kingdom\\
$^2$Mathematical Institute, University of Oxford\\ Woodstock Road, Oxford OX2\,6GG, United Kingdom\\
$^3$Oxford-Man Institute of Quantitative Finance\\
Walton Well Road, Oxford OX2\,6ED, United Kingdom
}

\begin{abstract}
\noindent 
We consider a rational agent who at time $0$ enters into a financial contract for which the payout is determined by a quantum measurement at some time $T>0$. 
The state of the quantum system is given in the Heisenberg representation by a known density matrix $\hat p$. How much will the agent be willing to pay at time $0$ to enter into such a contract? In the case of a finite dimensional Hilbert space 
$\mathcal H$, each such claim is represented by an observable 
$\hat X_T$ where the eigenvalues of $\hat X_T$ determine the amount paid if the corresponding outcome is obtained in the measurement. We prove, under reasonable axioms, that there exists a pricing state $\hat q$ which is equivalent to the physical state $\hat p$ such that the pricing function $\Pi_{0T}$ takes the linear form $\Pi_{0T}(\hat X_T) = P_{0T}\,{\rm tr} ( \hat q \hat X_T) $ for any claim $\hat X_T$, where $P_{0T}$ is the one-period discount factor. By ``equivalent" we mean that 
$\hat p$ and $\hat q$ share the same null space: thus, for any $|\xi \rangle \in \mathcal H$  one has $\langle \bar \xi | \hat p | \xi \rangle = 0$ if and only if 
$\langle \bar \xi | \hat q | \xi \rangle = 0$. We introduce a class of optimization problems and solve for the optimal contract payout structure for a claim based on a given measurement. Then we consider the implications of the Kochen-Specker theorem in this setting and we look at the problem of forming portfolios of such contracts. Finally, we consider multi-period contracts. 
\vspace{-0.2cm}
\\
\begin{center}
{\scriptsize {\bf Key words: Quantum mechanics, quantum measurement, contingent claims, discount bonds, \\ absence of arbitrage, rate of return, density matrices,  Gleason's theorem, Kochen-Specker theorem.
} }
\end{center}
\end{abstract}

\maketitle
\section {Introduction}
\label{sec:Introduction} 

\noindent By ``quantum finance" we mean the valuation, optimization and risk management of financial contracts for which the outcomes (in the form of one or more payments made between the various parties involved) are contingent on the results of one or more quantum measurements. The financial contracts that we consider can be easily implemented in a suitable laboratory. Our investigations fall within the scope of standard quantum mechanics and we are not concerned here with modifications of the standard framework or with interpretive issues. The resulting theory of quantum financial contracts is distinctly non-Kolmogorovian, inheriting as it does the full generality of quantum probability. 

The idea of forging connections between quantum theory and finance theory is not a new one. Previous attempts  have tended to fall into two broad categories. In the first category one has theories that work with the suggestion that asset prices -- and perhaps other economic variables as well -- are somehow subject to the laws of quantum mechanics.  

This is not an unreasonable thing to think about, given the rather general notions of ``complementarity" promoted by Bohr and Heisenberg. Nonetheless, it is probably safe to say that little by way of real progress has been made in this direction. The same is probably true if one tries to explain the overall macroeconomic characteristics of financial markets by use of statistical mechanics. For if asset prices were in some way subject to the laws of quantum mechanics, then surely the properties of large systems of prices could be explained by quantum statistical mechanics -- but that does not seem to be the case. 

In the second category of connections between quantum theory and finance, less controversially, one sees mathematical techniques originating in quantum theory being applied to problems in finance. Recently, we have witnessed a flurry of activity in the use of such techniques to improve on the traditional methods of computational finance, paving the way for the use of quantum computers to perform large-scale financial calculations. It should be emphasized, nonetheless, that no quantum ideas {\it per se} are involved in the finance theory underlying  such endeavours, so the term ``quantum finance" may be a misnomer in that context. See the Appendix for further discussion of previous approaches to quantum finance. 

Our theory represents a departure from these approaches. On the one hand, we are not suggesting the existence of quantum properties in ordinary financial assets such as stocks and bonds. Nor are we concerned with the use of ideas derived from quantum theory to speed up the risk management of conventional financial assets, or even new assets, such as those based on blockchain technology. In particular, we are not concerned with the use of quantum computers to tackle the work currently being carried out by classical computers. 

We are concerned, rather, with the pricing of securities for which the payouts are directly linked, by design, to the outcomes of quantum measurements. Needless to say, such financial products do not exist at present. But they may exist in the future, and that is why the theory is of interest from a scientific perspective. 
In particular, one can envisage the existence of a rather general type of structured product that delivers a sequence of cash flows, each being determined by the outcomes of a sequence of experiments. In such contracts, the experiments themselves may be adapted, in the sense that experiments performed at a later stage of the sequence may depend in their design on the outcomes of experiments made at an earlier stage of the sequence.  Here we consider the most basic of such structured products.

In Section \ref{sec:quantum measurements} we set out what we mean by a financial contract in a one-period market for which the payout is determined by a quantum measurement. We show that such a contract can be represented by a quantum observable (a Hermitian operator) for which the eigenvalues represent the possible cash flows. We begin with the example of a two-dimensional Hilbert space, for which the underlying experiment triggering the outcome of the contract involves measuring the spin of a spin $\half$ particle along a certain choice of axis. The state of the particle before the measurement is given by a known density matrix. The contract specifies the payments made for the two possible outcomes. The totality of these contracts constitute the ``market" associated with such spin measurements. 

In Section \ref{sec:Financial observables} we extend the discussion to the case of an $n$-dimensional Hilbert space $\mathcal H$ and we introduce the notion of a one-period discount bond, which pays out one unit of account at time $T$ regardless of the outcome of the experiment. In that case the associated financial observable is the identity operator on $\mathcal H$. 

We also introduce the notion of an Arrow-Debreu security, as it arises in the present context, for which the underlying experiment takes the form of a projection operator of rank unity. Such a contract either pays one unit of account or nothing, depending on which eigenvalue of the projection operator is attained when the measurement is performed.

The physical state of the underlying quantum system on which the measurement is taken is represented by a density matrix $\hat p$. We say that two density matrices 
$\hat p$ and $\hat q$ are {\it equivalent} if for all $|\psi \rangle \in \mathcal H$ it holds that $\hat p |\psi \rangle = 0$ if and only if $\hat q |\psi \rangle = 0$, i.e., they share the same null space. The significance of this equivalence relation among density matrices becomes apparent later when we consider the pricing of contracts. 
We argue that if two contracts 
$\hat U_T$ and $\hat V_T$ depend on the outcome of {\em the same experiment}, and differ from one another only in the amounts paid for the various outcomes of the experiment, then the prices of these contracts should satisfy a linear relation of the form
\begin{eqnarray}
\Pi_{0T} (a \hat U_T + b \hat V_T) = a \Pi_{0T} (\hat U_T) + b \Pi_{0T} (\hat V_T),
\end{eqnarray}
where $\Pi_{0T}$ denotes the pricing map. More precisely, if  
the operators $\hat U_T$ and $\hat V_T$ {\it commute}, then the prices should be additive. 

In Section \ref{sec:Existence of Pricing Operator} we present our main result, which is to show that under a certain set of axioms the pricing map necessarily takes the form 
\begin{eqnarray}
\Pi_{0T}\large[ \hat X_T\large] = P_{0T} \, {\rm tr} ( \hat q  \hat X_T ),
\end{eqnarray}
for some density matrix  $\hat q$, subject only to the condition that  $\hat p$ and $\hat q$ are equivalent.
The axioms are surprisingly simple. The first is that the price of a non-negative contract should vanish if and only if the expectation value of the contract vanishes. The second is that the pricing function should act linearly on any set of mutually commuting contracts. And the third is that the price of the observable corresponding to the identity operator should be that of a unit discount bond, which we regard as an input to the model. It should be emphasized that (a) we do not assume that the pricing map is linear, and that (b) no portfolio arguments are involved. The key ingredient in the proof is Gleason's theorem, which turns out to be surprisingly well adapted for applications in a financial context.

In Sections \ref{sec:Optimal Investment} and \ref{sec:Rate of Return} we look at a well-known classical investment problem in a quantum context. The problem is that faced by an investor with a fixed budget who wishes to invest optimally in such a way as to maximize expected utility. Like the classical problem, the quantum problem can be solved exactly, and although the mathematical ideas run in parallel in the two theories, it is not {\it a priori} obvious what form the solution of the quantum problem will take. 
In particular, if the quantum investment problem is generalized in such a way as to involve a choice between several incompatible experiments to determine the payout of the contract, then the problem cannot even be formulated in the classical theory of finance, and yet admits a neat formulation and solution in the case of quantum securities. 

This is shown in Section \ref{sec:Kolmogorov vs Bell-Kochen-Specker}, where we discuss more generally the role of quantum probability in finance. The example we consider is based on a construction of the Kochen-Specker type due to Cabello et al \cite{cabello et al, cabello 1997} 
involving a collection of nine incompatible observables. In this way we can formulate a quantum optimization problem, with no classical analogue, in which the investor faces a choice between nine different quantum financial contracts. 

In Section \ref{sec:Portfolios as Multi-Particle Systems} we return to the problem of portfolios, which we treat as structured products where the payouts depend on the results of two or more experiments. In the case of a one-period market, we consider the situation where one carries out measurements simultaneously on a pair of particles. The particles are associated with distinct Hilbert spaces, so no incompatibilities arise between the measurements and results are obtained for each. As a consequence, financial contracts can be devised for which payouts are made by totalling the results of each of the experiments. 

The two contracts can then consistently be regarded as part of the same portfolio in such a setup. The density matrix of the two-particle system as a whole can be entangled, allowing for correlations between the outputs of the individual constituents. A similar situation arises for portfolios involving any number of constituents. The surprising feature here is the relation between the ideas of entanglement in quantum mechanics and cointegration in portfolio theory. We conclude in Section \ref{sec:Conclusion} with a brief discussion of multi-period markets.

\section {Quantum measurements}
\label{sec:quantum measurements} 

\noindent Let time $0$ be the present and $T$ a fixed time in the future. We consider the situation where an agent $A$ enters into a contract with another agent $B$ in accordance with which $A$ pays $B$ an amount $H_0$ (``the price") at time $0$ and then $B$ pays $A$ an amount $H_T$ (the ``payout") at time $T$, where $H_T$ is contingent in some specified way on the outcome of  a  quantum measurement. We refer to such a setup as a one-period market.

By a quantum measurement, we mean the measurement of an observable associated with a microscopic system, such as a particle, or an atom or a molecule. 
More elaborate setups can be considered, involving multiple measurements, multiple payments and multiple agents; but for simplicity we look at a one-period market involving two agents. 
As an example, suppose the payout is determined by a measurement of the spin  of a spin one-half particle along the $z$-axis. The outcome of such a measurement either gives $+\half \hbar$, corresponding to spin up along that axis, or $-\half \hbar$, corresponding to spin down.
Henceforth, we work with physical units such that $\hbar = 1$. For the basics of quantum theory, see, for instance, reference \cite{isham}. We fix a two-dimensional Hilbert space $\mathcal H^2$ and on it we introduce the usual observable for the spin along the $z$-axis, given by the Hermitian operator
\begin{eqnarray}
\hat S_z = \half  |z_1\rangle \langle \bar z_1 | - \half |z_2\rangle \langle \bar z_2 |, 
\end{eqnarray}
where $|z_1\rangle$ is a unit Hilbert space vector corresponding to the upward direction along the $z$-axis and $|z_2\rangle$ denotes an orthogonal unit Hilbert space vector corresponding to the downward direction along the $z$-axis. Thus, $\langle \bar z_1 | z_1\rangle =1$, $\langle \bar z_2 | z_2\rangle =1$, $\langle \bar z_1 | z_2\rangle =0$,
$\langle \bar z_2 | z_1\rangle = 0$, and the possible outcomes of the measurement are the eigenvalues of 
$\hat S_z$, which are $+\half$ and $-\half$. 

The probabilities of these outcomes are determined by the {\it state} of the system, which is represented by a density matrix $\hat p$. The density matrix in quantum theory has a status that is analogous in certain respects to that of the probability measure in classical probability theory. The density matrix is assumed to be a positive-semidefinite Hermitian operator with trace unity, which in the case of a two-dimensional Hilbert space takes the form 
\begin{eqnarray}
\hat p = p_1  |\psi_1\rangle \langle \bar \psi_1 | + p_2 |\psi_2\rangle \langle \bar \psi_2 |, 
\end{eqnarray}
for some orthonormal basis  $\{  |\psi_1\rangle, |\psi_2\rangle \! \}$ in $\mathcal H^2$, where $p_1 \geq 0, \,p_2 \geq 0$, and $p_1 + p_2 = 1$. In general, such a matrix will have rank two, but if $p_1 = 0$ or $p_2 =0$ then it will have rank one. A state with rank one is called a ``pure" state. 

The probability for a given outcome is the trace of the product of the state and the projection operator onto the Hilbert subspace associated to the eigenvalue corresponding to that outcome (the ``Born rule"). Thus we have 
\begin{eqnarray}
{\rm Prob} \,\left( S_z = \half\right) =\langle \bar z_1 | \hat p |z_1\rangle, \quad
{\rm Prob} \,\left( S_z = -\half\right) =\langle \bar z_2 | \hat p |z_2\rangle.
\end{eqnarray}

In the case of a contingent claim where the payout is determined by the result of such a spin measurement, it should be clear that the claim itself can also be represented by a Hermitian operator on $\mathcal H^2$, in this case, an operator of the form
\begin{eqnarray} \label{financial observable}
\hat Z_T = z_1  |z_1\rangle \langle \bar z_1 | + z_2 |z_2\rangle \langle \bar z_2 |, 
\end{eqnarray}
where $z_1$ denotes the payment made to agent $A$ in the case the measurement outcome is spin $+\half$ and $z_2$ is the payment made to $A$ when the measurement outcome is spin $-\half$. One can think of such a contract as being an example of a so-called real option \cite{dixit pindyck, hughston zervos, trigeorgis}. Payments are understood to be made in some fixed numeraire or unit of account. Thus, we conclude that {\em a contingent claim for which the payouts are determined by the result of a quantum measurement can be represented by an observable, in the usual quantum mechanical sense, whose eigenvalues correspond to the possible cash flows at time $T$}. 

Among the various observables that can be represented in the form \eqref{financial observable}  there is a special observable that takes the form 
\begin{eqnarray} 
\hat P_{0T} = 1  |z_1\rangle \langle \bar z_1 | + 1 |z_2\rangle \langle \bar z_2 |,
\end{eqnarray}
which  pays one unit of account at time $T$, regardless of the outcome of the spin measurement. This is evidently a ``risk-free" asset, since the payout is fixed and guaranteed, and we write
\begin{eqnarray} \label{discount bond}
\hat P_{0T} =   \hat {\bf 1}.
\end{eqnarray}
Here $ \hat {\bf 1}$ denotes the identity operator on $\mathcal H^2$. The risk-free asset $\hat P_{0T}$ represents a discount bond that pays one unit of account (e.g., one ``dollar") at maturity $T$. It has the property that it does not depend on the choice of axis along which the spin measurement is taken.  

In addition to contracts of the form \eqref{financial observable}, we can more generally consider contracts of the same type, but where the measurement of the spin is taken along some other axis. Each such contract is characterized by (a) the choice of a basis in Hilbert space along which the spin measurement is made, together with (b) the payouts that take place as a consequence of the results of the measurement. Indeed, it is a theorem that any positive Hermitian operator $\hat Z_T$ on $\mathcal H^2$ other than multiples of the identity can expressed uniquely in the form \eqref{financial observable} for some choice of the orthonormal basis $\{  |z_1\rangle, | z_2\rangle \!\}$ in $\mathcal H^2$, modulo multiplicative phase factors.

To complete the discussion we need to determine the {\it price} paid by agent $A$ to agent $B$ in exchange for the payout corresponding to $\hat X_T$. In short, we need a {\it pricing function} that maps each financial observable $\hat X_T$ to a corresponding price $X_0$. It should be emphasized that there is nothing mysterious or obscure about the construction of such a market. One is left with the question of why an agent would wish to purchase a quantum security, but as in all economic considerations the issue of why there is supply and demand for a certain product is a matter quite distinct from the issue of how the market for that product will function, given that there is indeed supply and demand. 

\section {Financial observables}
\label{sec:Financial observables} 

\noindent It will be useful going forward to generalize our considerations to the case of a Hilbert space $\mathcal H$ of arbitrary finite dimension $n$. As usual, we can write $|\xi \rangle$ for a typical element of $\mathcal H$ and $\langle  \bar \xi |$ for its complex conjugate. The observable that determines the payout will in the generic situation be a non-degenerate Hermitian operator $\hat X_T$ on this space and hence admit $n$ distinct real eigenvalues, each corresponding to a distinct cash flow. 

For example, if the quantum system admits $n$ different energy levels, and the underlying physical observable being measured is the energy of the system, then the contract will in general result in a different cash flow 
$\{  x_j \}_{j = 1, 2,. \,. \,. \,, n}$ for each of the possible energy outcomes. For the financial observable representing such a contract we can write
\begin{eqnarray} \label{n dimensional financial observable}
\hat X_T = \sum_{j = 1}^n x_j  |x_j\rangle \langle  \bar x_j |, \quad \langle  \bar x_j |x_k\rangle = \delta_{jk}
\end{eqnarray}
for some orthonormal basis $\{  |x_j\rangle \!\}_{j = 1,2, . \,. \,. \,, n}$ in Hilbert space. More generally, the set of all financial observables associated with a given Hilbert space will include some that are degenerate in the sense that the same payout will result for two or more distinct values of the outcome $j$. Such a degeneracy can result either because there is a degeneracy in the spectrum of the underlying physical observable, or because two or more distinct eigenvalues of the physical observable are assigned the same cash flow. An example of the latter is a unit discount bond, for which $x_j = 1$ for all $j= 1,2, . \,. \,. \,, n$ even though the underlying energy levels may be distinct. Then the identity operator on 
$\mathcal H$ represents such a bond and again we write \eqref{discount bond} in that case. 

Another example of a degenerate observable is the analogue of a so-called Arrow-Debreu (A-D) security \cite{arrow debreu}, which for each value of $j$ has the payout $x_j = \mathds 1 \{ j = k\}$ for some fixed value of $k$. Here  $\mathds 1 \{ E\}$ denotes the indicator function for the event $E$. Thus  $x_j = 1$ if $j = k$, and $x_j = 0$  if $j \neq k$. The A-D securities are represented by pure projection operators, each with payout unity or zero, depending on the result of the underlying quantum measurement, whose outcome is also unity or zero. Thus the set of all Arrow-Debreu contracts is precisely the set of all pure projection operators on $\mathcal H$. 

The state of a quantum system in $n$ dimensions is represented by a positive semidefinite Hermitian matrix with trace unity. Such a matrix can be put in the form
\begin{eqnarray}
\hat p = \sum_{j = 1}^n p_j  |\psi_j\rangle \langle \bar \psi_j |, 
\end{eqnarray}
for some orthonormal basis  $\{  |\psi_j\rangle \!\}_{j = 1,2, . \,. \,. \,, n}$, with $p_j \geq 0$ for $j = 1,2, . \,. \,. \,, n $ and 
$\sum_{j = 1}^n p_j = 1$. 

In the case of a density matrix of maximal rank with distinct eigenvalues, this basis is uniquely determined up to phase factors. If the density matrix is of maximal rank but with a degenerate spectrum, the basis is determined modulo unitary transformations on the degenerate subspaces. In the case of a density matrix of lower rank, the basis is determined at best only up to an arbitrary unitary transformation of the basis vectors that span the null space of the density matrix. 

Given two density matrices  $\hat p$ and $\hat q$, we say that $\hat q$ is {\it absolutely continuous} with respect to $\hat p$ if the null space of $\hat p$ is a subspace of the null space of $\hat q$. Thus, $\hat q$ is absolutely continuous with respect to $\hat p$  if and only if for all $|\psi \rangle \in \mathcal H$ such that $\hat p |\psi \rangle = 0$ it holds that $\hat q |\psi \rangle = 0$. We say that $\hat p$ and $\hat q$ are {\it equivalent} if each is absolutely continuous with respect to the other, that is to say, if they share the same null space. 
%
%
It is easy to see that ``equivalence" in this sense  is an equivalence relation in the usual mathematical sense, and it follows that all density matrices of maximal rank are equivalent. 

We say that a claim $\hat X_T$ is {\it positive} if $\langle \bar \psi | \hat X_T   |\psi \rangle \geq 0$ for all 
$|\psi \rangle \in \mathcal H$ and {\it strictly positive} if $\langle \bar \psi | \hat X_T   |\psi \rangle > 0$ for all 
$|\psi \rangle \in \mathcal H$. By \eqref{n dimensional financial observable} one sees that  $\hat X_T$ is positive if and only if $x_j \geq 0$ for all $j$ and strictly positive if and only if $x_j > 0$ for all $j$. 

In fact, it should be evident that any claim $\hat X_T$ can be split in a canonically minimal way into positive part $\hat X^+_T$ and a negative part $\hat X^-_T$ such that $\hat X_T = \hat X^+_T + \hat X^-_T$, where the positive eigenvalues of $\hat X_T$ are those of $\hat X^+_T$ and the negative eigenvalues of $\hat X_T$ are those of $\hat X^-_T$. It will thus suffice for our purpose to look at financial contracts with positive cash flows

Let us now consider a one-period market represented by the set of all positive claims on an $n$-dimensional Hilbert space. The problem is to assign a value or price to each such claim $\hat X_T$ on the basis of as few assumptions as possible. One possible approach would be to consider the so-called expectation value of $\hat X_T$ in the state $\hat p$, given by 
\begin{eqnarray}
\langle \hat X_T \rangle_p = {\rm tr}( \hat p \hat X_T).
\end{eqnarray}
The expectation value can be interpreted as the average value of the payoff when the average is calculated by taking numerous independent copies of the experimental setup, performing an identical measurement on each system, and averaging the results. One might think that this expectation gives a fair price for entering into the contract; but that is merely a guess. In fact, agents will typically pay less than the expectation value, in order to allow for a non-trivial rate of return on the investment in compensation for the risks involved, and in principle, the price could be any non-negative map $\Pi_{0T}: \hat X_T \mapsto  \Pi_{0T}( \hat X_T) \in \mathbb R^+$, which need not necessarily be linear. 
On the other hand, we can be confident that if $\langle \hat X_T \rangle_p =0$, then the price must be zero, since no rational agent would pay a strictly positive premium for an investment that paid zero with probability one. Thus we conclude that the price vanishes if and only if the expectation value of the payoff vanishes. 

But we are still some distance from determining the form that the price takes. Since the expectation value is a linear function of the observable, this suggests that we look more closely at linear functionals. If $\hat X_T$ and $\hat Y_T$ are claims, then so is  the linear combination 
\begin{eqnarray} \label{linear combination}
\hat Z_T  = a \hat X_T + b \hat Y_T 
\end{eqnarray}
for $a,b \geq 0$. Hence the space of positive claims has a convex structure. It should be clear that the experiments underlying the $\hat X_T$ and $\hat Y_T$ are in general different and that the experiment underlying $\hat Z_T$ is different yet again.  If we write these claims in their diagonalized forms 
\begin{eqnarray}
\hat X_T = \sum_{j = 1}^n x_j  |x_j\rangle \langle \bar x_j |, \quad  \hat Y_T = \sum_{j = 1}^n y_j  |y_j\rangle \langle \bar y_j |, 
\end{eqnarray}
with respect to the relevant basis vectors, one sees that the payouts and basis vectors associated with these claims are uniquely determined, up to the usual ambiguities associated with degeneracies and null spaces, and at the same time the payouts and basis vectors of  \eqref{linear combination} are represented by the decomposition 
\begin{eqnarray}
\hat Z_T = \sum_{j = 1}^n z_j  |z_j\rangle \langle \bar z_j |.
\end{eqnarray}

Thus,  if we have two contracts, each with positive payouts, depending on separate measurements, then any linear combination of the operators corresponding to the two contracts, with positive coefficients, will give rise to the operator corresponding to yet another contract, with a different set of payouts, depending on still another measurement. 

Hence, a linear combination \eqref{linear combination} is {\it not}, generally, to be understood as representing a ``portfolio" of its constituents (see Section \ref{sec:Portfolios as Multi-Particle Systems}). This is because the payout of a portfolio is given by the totality of the payouts of its constituents. One is tempted, nonetheless, to conjecture that the price of the contract represented by a linear combination of two contracts should equal the corresponding linear combination of the prices of the constituents. But it is not obvious that this will be the case, since the new contract involves a different payout structure and a different experiment -- so we do not wish to {\em assume} linearity in general. 

We can, however, quite reasonably assume that such a linear relationship holds in certain special situations. In particular, if two contracts 
$\hat U_T$ and $\hat V_T$ depend on the outcome of {\em the same experiment}, and differ from one another only in the amounts paid for the various outcomes of the experiment, then the price of the contract $a \hat U_T + b \hat V_T$
should indeed be equal to the corresponding linear combination of the prices of $\hat U_T$ and $\hat V_T$. More precisely, if  
the $\hat U_T$ and $\hat V_T$ {\it commute}, then the prices should be additive. 
For if $\hat U_T$ and $\hat V_T$ commute, we can find a orthogonal basis $\{  |w_j\rangle \}_{j = 1,2, \, . \,. \,. \,, n}$ in which both are diagonalized: 
\begin{eqnarray}
\hat U_T = \sum_{j = 1}^n u_j  |w_j\rangle \langle \bar w_j |, \quad  \hat V_T = \sum_{j = 1}^n v_j  |w_j\rangle \langle \bar w_j |. 
\end{eqnarray}
Then if we form the linear combination $\hat W_T = a \hat U_T + b \hat V_T$ we obtain
\begin{eqnarray}
\hat W_T = \sum_{j = 1}^n (a u_j + b v_j)  |w_j\rangle \langle \bar w_j |, 
\end{eqnarray}
showing that the payouts for $\hat W_T$ are given by  linear combinations of the payouts of the constituents. Thus, {\it for commuting observables, the price of a linear combination of contracts should be the corresponding linear combination of the prices of the individual contracts}. But it is not obvious that linearity extends to non-commuting contracts. 

\section {Existence of Pricing Operator}
\label{sec:Existence of Pricing Operator} 
\noindent At this point, it may be helpful if we codify our assumptions somewhat more explicitly. 
As usual, we write $\mathbb R^+ = \{x \in \mathbb R : x \geq 0\}$. We fix a quantum system with state $\hat p$ on an $n$-dimensional Hilbert space  $\mathcal H$ and write $\mathcal V^+$ for the cone for positive contracts on $\mathcal H$. Thus our market is characterized by the triple $\{\mathcal H, \hat p,  \mathcal V^+\}$. Let us write $P_{0T}$ for the price of a unit discount bond.  Our goal is to assign a price to each contract $\hat X_T \in \mathcal V^+$. 
By a {\it pricing function} on the market $\{\mathcal H, \hat p, \mathcal V^+\}$ in a one-period setting we mean a mapping $\Pi_{0T}:  \mathcal V^+ \to \mathbb R^+ $ satisfying the following:
\vspace{0.25cm}

{\bf Axiom (1).} For all $\hat X_T \in \mathcal V^+$ it holds that $\Pi_{0T} [ \hat X_T] = 0$ if and only if ${\rm tr}( \hat p \hat X_T)$ = 0. 
\vspace{0.15cm}

{\bf Axiom (2).} If the $m$ contracts  represented by the Hermitian matrices $\{\hat X_T^k\}_{ k = 1, 2, \,. \, . \, .\, ,m}$ \newline \indent commute, then for all $\{a_k \geq 0\}_{ k = 1, 2, \,. \, . \, .\, ,m}$ one has
\begin{eqnarray}
\Pi_{0T}\! \left [ \sum_{k=1}^m a_k \, \hat X_T^k \right ] = \sum_{k=1}^m a_k \, \Pi_{0T} \!\left [ \hat X_T^k\right ]\!. 
\end{eqnarray}

{\bf Axiom (3).} $\Pi_{0T}[\, \hat {\bf 1}\,] = P_{0T}$.
\vspace{0.25cm}

\noindent The axioms can be interpreted as follows. Axiom (1) ensures the absence of arbitrage: the price of a positive contract  vanishes if and only if the expected payout vanishes.  Axiom (2) ensures that the pricing function is linear when it acts on a collection of contracts represented by commuting observables. Axiom (3) fixes the price of the risk-free asset. Then we obtain the following general characterization of the price of a contract:
\begin{Proposition}\label{prop:1}
If $n \geq 3$ then there exists a state $\hat q$ on  $\{\mathcal H, \hat p, \mathcal V^+\}$ that is equivalent to $\hat p$ such that for any contract $\hat X_T \in  \mathcal V^+$ the price of $\hat X_T$ is given by
\begin{eqnarray}\label{pricing formula}
\Pi_{0T}\large[ \hat X_T\large] = P_{0T} \, {\rm tr} ( \hat q  \hat X_T ).
\end{eqnarray}
\end{Proposition}
\noindent \textit{Proof}.
Consider the pricing of A-D securities. For each such contract, the measurement involves a projection operator $\hat \Lambda =  |\lambda \rangle \langle \bar \lambda | $
for some normalized vector $|\lambda \rangle \in \mathcal H$. The pricing function is a map from the space of pure projections on $\mathcal H$ to $\mathbb R^+$. It is well known that the space of pure projections on an $n$-dimensional Hilbert space is isomorphic to the complex projective space $\mathbb{CP}^{n-1}$. Thus we obtain a function 
$\Pi_{0T}: \mathbb {CP}^{n-1} \to \mathbb R^+$ with the property that for any $n$ points 
$\{\lambda_j \in \mathbb {CP}^{n-1} \} _{j = 1, 2, . \, . \, .\, ,n}$ 
determining an orthogonal basis in $\mathcal H$  one has
\begin{eqnarray}
\sum_{j = 1}^{n} \Pi_{0T}(\lambda_j) = P_{0T}.
\end{eqnarray}
This is because the projection operators associated with an orthonormal basis commute and hence by Axiom (2) the sum of the prices of the projection operators must equal the price of the sum of the projection operators. But the latter sum gives the identity operator, which offers a risk-free payout of unity. Thus we obtain a unit discount bond, for which the price is $P_{0T}$ by Axiom (3). 
Gleason's theorem \cite{gleason} can now be applied to the problem and it follows that there exists a state $\hat q$ such that the price of any claim of the form $\hat \Lambda$ is given by
\begin{eqnarray}
\Pi_{0T}\large[ \hat \Lambda \large] = P_{0T} \, {\rm tr} ( \hat q \hat \Lambda ).
\end{eqnarray}
Now, any contract $\hat X_T$ can be constructed as a linear combination of orthogonal pure projection operators
with positive coefficients. Since these operators commute, Axiom (2) implies that the price of such a contract will be given by the sum of the prices of its elements, and this gives us \eqref{pricing formula}. The fact that the ``pricing" operator $\hat q$ must be equivalent to the  ``physical" state $\hat p$ then follows as a consequence of Axiom (1), which taken with \eqref{pricing formula} ensures that for any positive contract $\hat X_T$ we have ${\rm tr} ( \hat p \hat X_T) = 0$ if and only if ${\rm tr} ( \hat q \hat X_T) = 0$. 
\hfill $\Box$
\vspace{0.5cm}

The point here is that we do not assume {\it a priori} the existence of a pricing state. The idea rather is to prove the existence of such a state under the {\it prima facie} much weaker assumptions implicit in our axioms. The requirement that the pricing function is linear when it is applied to any commuting family of A-D securities coupled with the assumption that the price of a one-period discount bond is known allows us to deduce that the pricing function 
takes the form \eqref{pricing formula}. In the case of a finite-dimension Hilbert space, the associated projective Hilbert space
 takes the form of a complex projective space $\mathbb{CP}^{n-1}$ equipped with the Fubini-Study metric  \cite{BH2001}. Gleason's theorem shows for $n \geq 3$ that any map $f: \mathbb{CP}^{n-1} \to [0,1]$ with the property 
that $\sum_{j=1}^n f(\lambda_j) = 1$ for any set of $n$ points $\{\lambda_j\}_{j = 1, 2, . \, . \, . \, n} \in \mathbb{CP}^{n-1}$ that are maximally distant from each other under the Fubini-Study metric  necessarily takes the form 
$f(\lambda) = \langle \bar \lambda | \hat q |\lambda \rangle   /  \langle \bar \lambda |\lambda \rangle $ for some positive operator $\hat q$ with trace unity. The principle of no arbitrage (``no free lunch") then implies that $\hat q$ is equivalent to $\hat p$. 

It should be noted that the physical state $\hat p$ refers to the state of the quantum system upon which measurement of a given physical observable determines the payment made under the terms of the financial contract. 
Thus $\hat p$ can be used to calculate the probability distribution of the payout, but gives no information about the price, except that minimal statement which is mandated by the absence of arbitrage -- namely, that the price should be zero if and only if the probability of a payout greater than zero is zero. 

The operator $P_{0T} \,\hat q$ plays the role of a pricing kernel in our theory.  In the case of an $n$-dimensional Hilbert space the prices of any $n^2 - 1$ linearly independent financial contracts, alongside the price of the unit discount bond, will be sufficient to completely calibrate the pricing kernel, which can then be used to price other contracts. It may seem surprising that the knowledge of such a system of prices gives no information about the physical state $\hat p$, except to determine its null space, but the analogue of this phenomenon is well established in the classical theory of finance \cite{bingham kiesel, dothan, etheridge, duffie, follmer schied}. At first glance, one might conclude that our Proposition 1 has little content, since the pricing operator $\hat q$ is arbitrary apart from its having the same null space as $\hat p$; but such a conclusion would be incorrect -- the point is that the existence of a pricing operator is not assumed but rather is {\em deduced} from the minimal axioms we have chosen to characterize a pricing function. Thus, beginning only with the assumed existence of a pricing function, which might in principle be nonlinear, one can whittle the candidates for such a map down to a linear function of the form \eqref{pricing formula}. 

\section {Optimal Investment}
\label{sec:Optimal Investment} 
\noindent A well-known problem in classical finance theory is to determine, given a budget $X_0$, the investment that maximizes the expectation of the utility gained by the investor at $T$ when the proceeds of the investment are liquidated.  It is reasonable to pose a similar problem in quantum finance. 
We assume that (a) agent $A$'s attitudes towards risk are expressed by a standard utility function  $\{U(x)\}_{x>0}$, (b) the physical state $\hat p$ of the quantum system  is known, (c) the basis under which the physical measurement is being made is known, and (d) the pricing state is known. The investment is thus characterized by an observable of the form 
\eqref{n dimensional financial observable}, where the basis $\{  |x_j\rangle \}_{j = 1, . \,. \,. \,, n}$ is fixed, and the cash flows $\{  x_j \}_{j = 1, . \,. \,. \,, n}$ must be determined so that the budget is saturated and the expected utility is maximized. 
What makes the problem interesting is that the expected utility of the payout is calculated by use of the physical state $\hat p$ whereas the budget constraint involves the pricing state $\hat q$, and that neither $\hat p$ nor $\hat q$ necessarily has any special relation to the measurement basis. 

\begin{Definition}\label{definition:1}
By a standard utility function we mean a map $U: \mathbb R^+\backslash \{0\} \to \mathbb R$ that satisfies the following conditions:  {\rm (i)} $U \in {\rm C}^2(\mathbb R^+\backslash \{0\})$, {\rm(ii)} $U'(x) > 0$ for all $x >0$, {\rm(iii)} 
$U''(x)< 0$ for all $x >0$, {\rm(iv)} $\lim_{x \to \infty} U'(x) = 0$, and {\rm(v)} $\lim_{x \to 0 } U'(x) = \infty$. 
\end{Definition}

These requirements can be relaxed in various contexts, but the ``standard" conditions often lead to well-posed problems for which solutions can be shown to exist and hence prove to be natural as a basis for modelling.  We see that a standard utility function is a strictly convex, strictly increasing map defined for all strictly positive values of its argument. 
We refer to the map
$U': \mathbb R^+\backslash \{0\} \to  \mathbb R^+\backslash \{0\}$ as the {\it marginal utility}. 
The final two conditions of the definition ensure that there exists an inverse marginal utility function $\{I(y)\}_{y>0}$ such that $I(U'(x)) = x$ for all $x>0$. The identity
\begin{eqnarray}
U(I(y)) - I(y) y = \sup_{x>0} \left( U(x) - xy\right),
\end{eqnarray}
which holds for all $y>0$, can be used to establish the so-called fundamental inequality
\begin{eqnarray} \label{fundamental inequality}
U(I(y)) - I(y) y \geq U(z) - yz,
\end{eqnarray}
which holds for all $y> 0$ and $z>0$ in the case of a standard utility function.

Examples of standard utility functions are (a) logarithmic utility, for which $U(x) = \log (x)$ for $x > 0$, and (b) power utility with index $p \in (-\infty, 1)\backslash \{0\}$, for which $U(x) = p^{-1}x^p$ for $x>0$. For logarithmic utility one finds that $I(y) = 1/y$ and for power utility  $I(y) = y^{1/(p-1)}$. 

The goal of agent $A$'s optimization problem is  to determine the cash flows $\{  x_j \}_{j = 1, 2, . \,. \,. \,, n}$ that maximize the expected value of the utility, providing that these cash flows can be realized with the specified budget.  Thus, given a standard utility function $\{U(x)\}_{x>0}$ we set
\begin{eqnarray}
\{  x^*_j \}_{j = 1, 2, \,. \,. \,. \,, n} = \argmax_{\{x_j\}} \,{\rm tr} \left[ \hat p \, \hat U(\{x_j\}) \right]
\end{eqnarray}
where
$\hat U(\{x_j\}) = \sum_{j = 1}^n U(x_j)  |x_j\rangle \langle \bar x_j |
$
and the argmax is subject to the budget constraint
\begin{eqnarray} \label{budget constraint}
X_0 = P_{0T} \, {\rm tr} ( \hat q  \,\hat X_T ), \quad \hat X_T = \sum_{j = 1}^n x_j  |x_j\rangle \langle  \bar x_j |. 
\end{eqnarray}
%

\begin{Proposition}\label{prop:2}
Let the physical state of a quantum system on an $n$-dimensional Hilbert space be $\hat p$. Let the pricing state for a financial market based on measurements of the system be $\hat q$, with one-period discount factor $P_{0T}$. Let the risk preferences of the investor be represented by a standard utility function $U: \mathbb R^+\backslash \{0\} \to \mathbb R$ and write 
$I$ for the associated inverse marginal utility function. Then the optimal cash flow structure $\{  x^*_j \}$ for an investment with budget $X_0$ paying out according to the measurement of a financial observable of the form
\begin{eqnarray}
\hat X = \sum_{j = 1}^n x_j  |x_j\rangle \langle \bar x_j | ,
\end{eqnarray}
for some fixed orthonormal basis $\{  |x_j\rangle \!\}_{j = 1, . \,. \,. \,, n}$, is given by
\begin{eqnarray} \label{optimal payouts} 
x^*_j = I \!\left [  \lambda \,P_{0T} \frac{ \langle \bar x_j | \hat q |x_j\rangle } { \langle \bar  x_j | \hat p | x_j\rangle}  \right]\!, 
\end{eqnarray}
where for any choice of $X_0> 0$ the parameter $\lambda$ is uniquely determined by the relation
\begin{eqnarray} \label{finalized constraint}
P_{0T}  \sum_{j = 1}^n  I \!\left [  \lambda \,P_{0T} \frac{ \langle \bar x_j | \hat q |x_j\rangle } { \langle \bar  x_j | \hat p | x_j\rangle}  \right]
 \langle x_j | \hat q |\bar x_j\rangle = X_0.
\end{eqnarray}
\end{Proposition}
\noindent \textit{Proof}. The method of Lagrange multipliers can be used to obtain a candidate for the argmax.
We introduce a Lagrange multiplier $\lambda$ and seek a solution to the unconstrained problem
\begin{eqnarray}
\{  x^*_j \} = \argmax_{\{x_j\}} \,\left ( {\rm tr} \left[ \hat p \, \hat U(\{x_j\}) \right] - \lambda P_{0T} \, {\rm tr} ( \hat q  \,\hat X_T ) \right )\!,
\end{eqnarray}
or equivalently 
\vspace{-0.25cm}
\begin{eqnarray}
\{  x^*_j \} = \argmax_{\{x_j\}} \, \left ( \sum_{j = 1}^n U(x_j) \langle \bar x_j | \hat p | x_j\rangle - 
\lambda P_{0T} \sum_{j = 1}^n x_j \langle \bar  x_j | \hat q | x_j\rangle \right )\!.
\end{eqnarray}
Differentiating with respect to $x_j$ and setting the results to zero, we find that
\begin{eqnarray}
U'(x_j) = \lambda \,P_{0T} \frac{ \langle \bar x_j | \hat q |x_j\rangle } { \langle \bar  x_j | \hat p | x_j\rangle}
\end{eqnarray}
for each value of $j$. Applying the inverse marginal utility function to each side of this equation, we are then led to
\eqref{optimal payouts} and the budget constraint \eqref{budget constraint} gives \eqref{finalized constraint}. That \eqref{finalized constraint} admits a unique solution for $\lambda$ for any $X_0 > 0$  follows 
from the fact that the monotonic decreasing map $I: \mathbb R^+\backslash \{0\} \to  \mathbb R^+\backslash \{0\}$ is surjective, which is a consequence of the conditions (iv) and (v) satisfied by a standard utility function.  That the candidate solution is indeed a true solution can be checked by use of the  fundamental inequality \eqref{fundamental inequality}.  It follows then from \eqref{optimal payouts} that  for any {\em alternative} choice of payout structure $\{  x_j \}$ we have
\begin{eqnarray}
U(x^*_j) - x^*_j  \lambda \,P_{0T}\frac{ \langle \bar x_j | \hat q |x_j\rangle } { \langle \bar  x_j | \hat p | x_j\rangle}  \geq U(x_j) - 
x_j \lambda \,P_{0T}\frac{ \langle \bar x_j | \hat q |x_j\rangle } { \langle \bar  x_j | \hat p | x_j\rangle},
\end{eqnarray}
 for each $j = 1, 2, . \,. \,. \,, n.$ Multiplying by $p_j$ and summing we obtain
\begin{eqnarray}
\sum_{j = 1}^n p_j U(x^*_j) - \sum_{j = 1}^n p_j U(x_j)   \geq  
\sum_{j = 1}^n p_j x^*_j  \lambda \,P_{0T}\frac{ \langle \bar x_j | \hat q |x_j\rangle } { \langle \bar  x_j | \hat p | x_j\rangle} - 
\sum_{j = 1}^n p_j x_j \lambda \,P_{0T} \frac{ \langle \bar x_j | \hat q |x_j\rangle } { \langle \bar  x_j | \hat p | x_j\rangle}.
\end{eqnarray}
Then since $p_j = \langle x_j | \hat p |\bar x_j \rangle$ we have 
\begin{eqnarray}\label{step in argument}
\sum_{j = 1}^n p_j U(x^*_j) - \sum_{j = 1}^n p_j U(x_j)   \geq  
\lambda \,P_{0T} \left [ \sum_{j = 1}^n  x^*_j   \langle \bar x_j | \hat q | x_j\rangle  - 
\sum_{j = 1}^n x_j  \langle \bar x_j | \hat q | x_j\rangle \right ]\!.
\end{eqnarray}
Now, we know by \eqref{finalized constraint} that $\lambda$ has been chosen to ensure that the candidate solution $\{  x^*_j \}$ satisfies the budget constraint 
\begin{eqnarray} \label{budget constraint formula}
P_{0T}  \sum_{j = 1}^n   x^*_j 
 \langle \bar  x_j | \hat q | x_j\rangle = X_0.
\end{eqnarray}
If we require that the alternative choice of payout structure should also satisfy the budget constraint, or else operate under budget, so 
\begin{eqnarray} 
P_{0T}  \sum_{j = 1}^n   x_j 
 \langle \bar x_j | \hat q | x_j\rangle \leq X_0,
\end{eqnarray}
then the two terms on the right-hand side of \eqref{step in argument} cancel, or else leave a difference that is positive (if the alternative choice is under budget), which gives
\begin{eqnarray} 
\sum_{j = 1}^n p_j U(x^*_j)  \geq   \sum_{j = 1}^n p_j U(x_j),  
\end{eqnarray}
showing that the candidate solution for the optimal payout gives an expected utility that is no less than that of any alternative choice of payout structure with a budget no greater than that of the candidate solution.
\hfill $\Box$
\vspace{0.5cm}
\section {Rate of Return}
\label{sec:Rate of Return} 

\noindent As an example, we can look in detail at the case of logarithmic utility. Suppose we set $U(x) = \log x$ for $x>0$. Then the inverse marginal utility function is given by $I(y) = 1/y$ for $y>0$. It follows that for log utility the optimal payout structure takes the form
\begin{eqnarray}
x^*_j =   ( \lambda \, P_{0T})^{-1} \frac{ \langle \bar x_j | \hat p|x_j\rangle } { \langle \bar  x_j | \hat q | x_j\rangle}. 
\end{eqnarray}
Inserting this expression into the budget constraint \eqref{budget constraint formula} we obtain
\begin{eqnarray} 
 \lambda^{-1} \sum_{j = 1}^n   
 \langle \bar  x_j | \hat p | x_j\rangle = X_0.
\end{eqnarray}
But the sum appearing in the expression above is unity since 
$ \sum_{j = 1}^n |x_j\rangle \langle  \bar x_j | = \hat {\bf 1}$
and the trace of $\hat p$ is one. Thus for log utility we deduce that $  \lambda^{-1} = X_0$ and hence
\begin{eqnarray} \label{x-star}
x^*_j =(P_{0T})^{-1} X_0 \frac{ \langle \bar x_j | \hat p|x_j\rangle } { \langle \bar  x_j | \hat q | x_j\rangle}. 
\end{eqnarray}

We observe that when the physical state and the pricing state are one and the same, the payouts of the optimal investment are identical for each outcome of chance, each giving  $(P_{0T})^{-1} X_0$, the usual ``future value" of the initial investment.
In that case,  the optimal investment is to put  the initial endowment into unit discount bonds, totalling $X_0$ in value. Then we have $\hat X_T =(P_{0T})^{-1} X_0 \, \hat {\bf 1}$. It follows that if the pricing state is the physical state, the market assigns no premium to the return on a risky investment, ensuring that the optimal investment is in a discount bond and the rate of return is the interest rate. 

The same conclusion applies, more generally, for any choice of the utility. This follows from \eqref{optimal payouts} and \eqref{finalized constraint}, from which one concludes that if $\hat p = \hat q$ then $x^*_j = (P_{0T})^{-1} X_0$ for all $j$. 
It is interesting therefore to enquire what happens when the pricing state is different from the physical state. The return 
$R_{0T}$ on an investment $\hat X_T$ is given by the ratio of the expectation of $\hat X_T$ under $\hat p$ to the amount initially invested, namely $X_0$. Thus, quite generally, we have 
\begin{eqnarray} 
R_{0T} = (X_0)^{-1}  {\rm tr} (\hat p \hat X_T).
\end{eqnarray}
But $X_0 = P_{0T}\,{\rm tr} (\hat q \hat X_T)$ by \eqref{pricing formula}, so we deduce that
\begin{eqnarray} 
R_{0T} = (P_{0T})^{-1}  \frac   { {\rm tr} (\hat p \hat X_T) } { {\rm tr} (\hat q \hat X_T) },
\end{eqnarray}
and it should be clear that if $\hat p = \hat q$, except possibly on the null space of $\hat X_T$, then the rate of return on the investment is the one-period interest rate.

Specializing now to the case of an optimal investment for an agent with logarithmic utility, let us calculate the rate of return. We have 
\begin{eqnarray}
\hat X_T = \sum_{j = 1}^n x^*_j \, |x_j\rangle \langle \bar x_j | ,
\end{eqnarray}
where the optimal payout structure $\{x^*_j\}$ is given by \eqref{x-star}. It follows then that
\begin{eqnarray} 
R_{0T} &=& (X_0)^{-1}  {\rm tr} (\hat p \hat X_T) \nonumber \\
&=&(X_0)^{-1} \sum_{j = 1}^n x^*_j \, \langle \bar x_j | \hat p|x_j\rangle \nonumber \\
&=& (P_{0T})^{-1} \sum_{j = 1}^n \frac{ \,\,\langle \bar x_j | \hat p|x_j\rangle^2 } { \langle \bar  x_j | \hat q | x_j\rangle}.
\end{eqnarray}
If we set $R_{0T} =\re^{\mu T}$ then the rate of return $\mu$ can be split into two parts, namely a risk-free one-period interest rate and a so-called excess rate of return or risk premium, which is the part of the rate of return that exceeds the interest rate. We can represent this by writing
$R_{0T} = \re^{(r + \beta) T}$ where $r$ is the interest rate and $\beta$ is the excess rate of return. The interest rate is fixed by the relation $\re^ {rT} = (P_{0T})^{-1}$ and the excess rate of return is fixed by the relation
\begin{eqnarray} \label{excess rate of return}
\re^ { \beta T} =  \sum_{j = 1}^n \frac{\,\, \langle \bar x_j | \hat p|x_j\rangle^2 } { \langle \bar  x_j | \hat q | x_j\rangle}.
\end{eqnarray}
%
\begin{Proposition}\label{prop:3}
The optimal investment in the case of an investor with logarithmic utility has a positive excess rate of return. The utility gained from such an investment in a market where the physical state and pricing state differ is greater than or equal to the utility gained from an investment in a risk-free bond.
\end{Proposition}
\noindent \textit{Proof}. 
The expected utility gained from the payout of an optimal investment is 
\begin{eqnarray} 
{\rm tr} \left[ \hat p \, \hat U\right] = \sum_{j = 1}^n U(x^*_j) \langle \bar x_j | \hat p | x_j\rangle.
\end{eqnarray}
Let us set $U(x^*_j) = \log x^*_j$ for logarithmic utility and insert \eqref{x-star}. 
The result is 
\begin{eqnarray} \label{expected utility}
\sum_{j = 1}^n U(x^*_j) \langle \bar x_j | \hat p | x_j\rangle = \log \left [(P_{0T})^{-1} X_0\right ] + \sum_{j = 1}^n \left[
\langle \bar x_j | \hat p | x_j\rangle \log \frac{ \langle \bar x_j | \hat p|x_j\rangle } { \langle \bar  x_j | \hat q | x_j\rangle} \right]\!.
\end{eqnarray}
The first term on the right-hand side of this equation isolates the part of the utility gain due to the interest rate. The second term can be interpreted as a {\it relative entropy}. In particular, if we set $p_j =  \langle \bar x_j | \hat p|x_j\rangle$ and  $q_j =  \langle \bar x_j | \hat q|x_j\rangle$  then it is evident that $\{p_j\}_{j = 1, 2, . \,. \,. \,, n}$ and $\{q_j\}_{j = 1, 2, . \,. \,. \,, n}$ constitute a pair of absolutely continuous probability distributions. The second term on the right then takes the form of a Kullback-Liebler divergence \cite{Kullback Leibler 1951}:
\begin{eqnarray} 
D_{KL}(p, q) = \sum_{j = 1}^n p_j \log \left(\frac {p_j} {q_j}\right)\!. 
\end{eqnarray}
Thus, the utility thus gained gives a measure of the divergence between the physical state and the pricing state. 
Now, it is well known that the Kullback-Liebler divergence is {\em non-negative}. 
It follows, then, that the utility gained from an optimal risky investment in a market where $\hat p$ and $\hat q$ are distinct will be greater than or equal to the utility gained from a risk-free bond investment, as claimed. 

Moreover, we have the following. The standard logarithmic inequality $\log z \leq z - 1$, which holds for $z > 0$, implies that
\begin{eqnarray} 
\log \left(\frac {p_j} {q_j}\right)  \leq \left(\frac {p_j} {q_j}\right) - 1
\end{eqnarray}
for each $j$. Hence, multiplying by $p_j$ and summing we obtain 
\begin{eqnarray} 
\sum_{j = 1}^n  p_j \log \left(\frac {p_j} {q_j}\right) \leq \sum_{j = 1}^n \left(\frac {p_j^2} {q_j}\right) - 1.
\end{eqnarray}
Thus, we have
\begin{eqnarray} 
\re^ { \beta T}  = \sum_{j = 1}^n \left(\frac {p_j^2} {q_j}\right) \geq 1 + D_{KL}(p, q), 
\end{eqnarray}
and by the positivity of the Kullback-Liebler divergence we deduce that the excess rate of return $\beta$ is positive for an optimal investment under logarithmic utility, as claimed. 
\hfill $\Box$
\vspace{0.5cm}
\section {Classical vs Quantum Probability}
\label{sec:Kolmogorov vs Bell-Kochen-Specker} 

\noindent It is often maintained that quantum probability is more general than Kolmogorov's well-established ``classical" theory of probability \cite{kolmogorov1933} and that the latter is contained as a special case of the former. There is no doubt that quantum probability, when laid out as a mathematical theory, has a different look and feel when it is compared to Kolmogorov's theory; but despite the fact that numerous well-argued accounts of quantum probability can be found in the literature  \cite{davies, deutsch, gudder, holevo, kraus, mackey, segal1947, streater2000} (see also \cite{BCFFS, deutsch, wallace 2010}), some even taking an axiomatic approach, it is not that easy to pinpoint the exact sense in which quantum theory is {\it essentially} non-Kolmogorovian -- rather than, say, a  reworking of Kolmogorov's theory in a different form. 
This issue is compounded by the fact that, except in the most loose terms, it is difficult to say what one means by ``probability" without embedding the concept in a mathematical framework.

It is fortunate then that we have the results of Gleason \cite{gleason}, Bell \cite{bell1964, bell1966, bell1987}, Kochen \& Specker \cite{kochen-specker, held}, and others following in their footsteps, which add clarity to the matter. The point is that one has to work rather hard to come up with examples of situations in quantum probability that cannot be reduced to a classical probability model. But a number of such examples have been worked out involving finite-dimensional Hilbert spaces, so this creates the prospect of constructing financial models for claims based on the results of quantum measurements, in settings for which quantum probability is required in their analysis. 
Since most of what we know of modern finance theory is based explicitly on Kolmogorov's framework, it may be worthwhile to take note of a few examples of situations where quantum probability comes into play. 

Among the numerous attempts that have been made to generalize or extend the Kochen-Specker construction  \cite{mermin, peres, kernaghan, kernaghan-peres, penrose}, perhaps the simplest yet put forward  is that of Cabello et al \cite{cabello et al, cabello 1997}, which entails the specification of a collection of nine different non-commuting observables  on a four-dimensional Hilbert space.

In a financial context, one can think of this setup as involving a single quantum system being prepared in a state $\hat p$ with nine different ``draft financial contracts" drawn up, each requiring measurement of one of the nine observables. The contracts specify the payments that will be made when one of the four possible outcomes occurs for the measurement associated with a specific contract. It is of the nature of quantum probability that only one of the nine contracts can be implemented, so we can envisage a rational agent being presented with the alternatives and choosing one optimally in accordance with their needs. 

In any specific setting, there will only be one contract in play, namely the one chosen by the agent after careful consideration of their criteria for optimality. In each such specific setting the usual rules for Kolmorovian probability apply. But for the setup and description of the problem as a whole -- with the presentation and analysis of the nine contracts and the posing of the optimization problem, we require quantum probability. 
This example can be used to refute the claim of a skeptic who asks whether one is merely taking simple examples from classical finance and dressing them up in the language of quantum probability and calling the result quantum finance. 
The point is that completely tractable examples can be constructed within the context of quantum finance for which no classical analogue exists. 

\begin{figure}[H]
    \centering
\begin{tikzpicture}
\newcommand{\numnodes}{9}
\newcommand{\numntriangles}{3}
\foreach \ii in {1, ..., \numntriangles}{
    \pgfmathsetmacro{\angleone}{90 - (\ii+2*(\ii-1)-2)*360/\numnodes}
    \pgfmathsetmacro{\angletwo}{90 - (\ii+2*(\ii-1)-1)*360/\numnodes}
    \pgfmathsetmacro{\anglethree}{90 -(\ii+2*(\ii-1))*360/\numnodes}
    \path[draw, thin, dashed] (\angleone:3cm) -- (\angletwo:3cm);
    \path[draw, thin, dashed] (\angletwo:3cm) -- (\anglethree:3cm);
    \path[draw, thin, dashed] (\anglethree:3cm) -- (\angleone:3cm);
}
\foreach \ii in {1, ..., \numntriangles}{
\foreach \ij in {1, ..., \numntriangles}{
    \pgfmathsetmacro{\startangle}{90 - (\ii+3*(\ij-1)-1)*360/\numnodes}
    \pgfmathsetmacro{\endangle}{90 - (\ii+3*(\ij)-1)*360/\numnodes}
    \path[draw, thin, dashed] (\startangle:3cm) -- (\endangle:3cm);
}
}

{
\pgfmathsetmacro{\angle}{360/\numnodes}
\pgfmathsetmacro{\zangle}{90-\angle*0}
\pgfmathsetmacro{\aangle}{90-\angle*1}
\pgfmathsetmacro{\bangle}{90-\angle*2}
\pgfmathsetmacro{\cangle}{90-\angle*3}
\pgfmathsetmacro{\dangle}{90-\angle*4}
\pgfmathsetmacro{\eangle}{90-\angle*5}
\pgfmathsetmacro{\fangle}{90-\angle*6}
\pgfmathsetmacro{\gangle}{90-\angle*7}
\pgfmathsetmacro{\hangle}{90-\angle*8}

\node[align=left, font=\scriptsize] at (\zangle:4.1cm) { 0 0 0 1 \\  0 1 0 0 \\  1 0 1 0 \\  1 0 $\bar 1$ 0}; 
\node[align=left, font=\scriptsize] at (\aangle:4.1cm) { 0 0 1 0 \\  0 1 0 0 \\  1 0 0 1 \\  1 0 0 $\bar 1$}; 
\node[align=left, font=\scriptsize] at (\bangle:4.1cm) { 1 $\bar 1$ 1 $\bar 1$ \\  1 $\bar 1$ $\bar 1$ 1 \\  1 1 0 0 \\  0 0 1 1};
\node[align=left, font=\scriptsize] at (\cangle:4.1cm) { 1 $\bar 1$ 1 $\bar 1$ \\  1 1 1 1 \\  1 0 $\bar 1$ 0 \\  0 1 0 $\bar 1$};
\node[align=left, font=\scriptsize] at (\dangle:4.1cm) { 1 $\bar 1$ $\bar 1$ 1 \\  1 1 1 1 \\  1 0 0 $\bar 1$ \\  0 1 $\bar 1$ 0};
\node[align=left, font=\scriptsize] at (\eangle:4.1cm) { 1 1 $\bar 1$ 1 \\  1 1 1 $\bar 1$ \\  1 $\bar 1$ 0 0 \\  0 0 1 1};
\node[align=left, font=\scriptsize] at (\fangle:4.1cm) { 1 1 $\bar 1$ 1 \\  $\bar 1$ 1 1 1 \\  1 0 1 0 \\  0 1 0 $\bar 1$};
\node[align=left, font=\scriptsize] at (\gangle:4.1cm) { 1 1 1 $\bar 1$ \\  $\bar 1$ 1 1 1 \\  1 0 0 1 \\  0 1 $\bar 1$ 0};
\node[align=left, font=\scriptsize] at (\hangle:4.1cm) { 0 0 0 1 \\  0 0 1 0 \\  1 1 0 0 \\  1 $\bar 1$ 0 0};
}
\foreach \ii in {1, ..., \numnodes}{
    \pgfmathsetmacro{\angle}{90 - (\ii-1)*360/\numnodes}
    \node[circle, draw, fill=blue!5] (x\ii) at (\angle:3cm) {};
}
\end{tikzpicture}

    \caption{Diagram illustrating a  result of the Kochen-Specker type in a 4-dimensional Hilbert space. Each of the 9 vertices are met by 4 lines and each of the 18 lines join 2 vertices.  The 18 lines represent a set of normalized projection operators with the property  that the 4 projection operators meeting a given vertex are mutually orthogonal and sum to the identity operator. It is easy to see that it is impossible to ``colour" the lines so that one blue line meets each vertex and 3 red lines meet each vertex.
This illustrates the fact that in the standard Kolmogorov setup one cannot find a set of 18 random variables on a probability space $(\Omega, \mathcal F, \mathbb P)$, each taking  values in the set $\{0, 1\}$, such that when the 18 random variables are assigned to the 18 lines, the sum of the 4 random variables meeting any given vertex will be one for all $\omega \in \Omega$. 
 }
    \label{fig:my_label}
\end{figure}
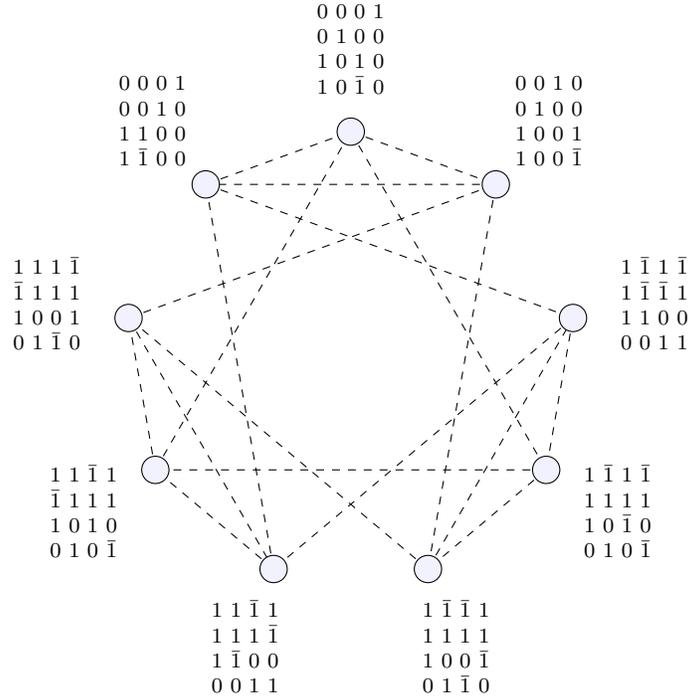

 The setup is an elaborate although feasible one, and we can use the methods discussed to calculate the probabilities of the results for the nine different measurements and hence the expected utility gained from each choice. 
Each observable has four possible outcomes, thus determining an orthonormal tetrad in Hilbert space. These are the four eigenvectors of the Hermitian matrix corresponding to a given observable. The result of the measurement is to select one of these eigenvectors. Equivalently, each measurement measures four commuting projection operators, namely the projection operators associated with the four legs of the tetrad. The outcome of one of these four measurements will be unity and the rest nil. 

The clever idea behind results of the Kochen-Specker type is to choose the observables so that some of the tetrads legs overlap when one moves from one observable to another. In the present situation, involving nine observables, the overlap structure is shown in Figure 1. Alongside each vertex of the enneagon one sees the corresponding tetrad, where to ease the typography we write $\bar 1$ for $-1$. When two vertices are connected by a dotted line, this means that the associated tetrads share a vector in common. The analysis is simplified somewhat by the fact that the tetrads in this example can all be taken to be real. 

If we label the nine observables $\{ \hat X_r \}_{r = 1, 2, \, . \, .\, . , 9}$ and if for each value of $r$ the four projection operators associated with $\hat X_r$ are denoted $\{\hat \pi_{rj}\}_{j= 1, 2, 3, 4}$\,, then the probability that outcome $j$ will result, if contract $r$ is chosen, is given by ${\rm tr}( \hat \pi_{rj}\, \hat p)$. The construction of an analogous setup within Kolmogorov's system turns out to be impossible. Since this is a rather sweeping statement, let us be a little more precise about what is being claimed. The point is that in Kolmogorov's theory, one would have to model the setup with 36 random variables on a single probability space. The 36 random variables are grouped into nine sets of four. Let's call these hypothetical random variables  $\{ X_{rj} \}_{r = 1, 2, \, . \, .\, . ,9, \, j= 1, 2, 3, 4}$ (with no hats). Each random variable can take the value zero or one. Thus we have a total of 36 maps
of the form 
\begin{eqnarray} 
X_{rj} : \Omega \to \{0, 1\}.
\end{eqnarray}
There are two requirements that have to be satisfied to match the layout of the quantum setup. First, the sum of the four random variables for a given value of $r$ must be unity. This means that one of them must be equal to one and the other three must be equal to zero for any given outcome of chance $\omega \in \Omega$. Secondly (this is where the rabbit goes into the hat) the 36 random variables have to be equal in pairs, in conformation with the structure of the diagram in Figure 1. Thus, the 36 random variables are cut down in effect to 18 by the requirement that they must match in pairs. 

Can one find such a set of 18 random variables? The answer, perhaps surprisingly, is no. This can be checked by a colouring argument. Given Figure 1, can one colour each line red or blue in such a way that exactly one blue line meets each vertex? Suppose one finds a way of colouring four of the lines blue, no vertex being hit by more than one blue line. That would leave one vertex unmet by a blue line. Suppose then one tried to colour five lines blue. Well, that would mean at least one vertex was hit by more than one blue line. This shows that it is impossible to construct a set of 18 random variables on a probability space in such a way that the required properties are satisfied. 

In financial terms, this means that we cannot model the payouts of the nine contracts as random variables on a probability space in such a way that the outcome of chance determines the payouts of all nine. A sceptic might ask, ``Isn't it unlikely in practice that one will come up against such a configuration of contracts?" Well, that may be so, but the point is that quantum finance can handle such configurations whereas classical finance can't. 

\section {Portfolios}
\label{sec:Portfolios as Multi-Particle Systems} 

\noindent Let us return now to the matter of portfolios. There are two rather distinct notions of portfolio that arise in quantum finance. The first notion involves a portfolio of contracts all depending in their payouts on the same experiment. In that case, we can fix the $n$ axes of the $n$-dimensional Hilbert space determining the frame of the measurement and write $\{\hat \pi_{j}\}_{j= 1, 2, \, . \, . \, . \, n}$ for the associated projection operators. 
Then, for a given outcome of the experiment one of these projection operators will give the result unity and the rest zero. The projection operators can be regarded as the A-D securities for that experiment and it should be evident that any contingent claim based on the outcome of the given experiment can be written as a portfolio of $n$ such A-D securities. Thus, for such claims we can write 
\begin{eqnarray} 
\hat X =  \sum_{j = 1}^n \theta_j \,\hat \pi_{j},
\end{eqnarray}
where the $\{ \theta_j\}_{j= 1, 2, \, . \, . \, . \, n}$ represent the holdings in the various A-D securities. More generally, if we allow short positions in the A-D securities, then the resulting overall position can be expressed uniquely as the difference between two positive claims, with the understanding that we net claims involving long and short positions in the same A-D security. 

Clearly, a linear combination of two portfolios in this setting gives another portfolio. Furthermore, it should be evident that the operator corresponding to the portfolio can be represented as the sum of a trace part, proportional to the identity operator, and a trace-free part. 
The trace part represents a position (long or short) in the risk-free asset, and the remainder consists of investments in risky assets. For example, in two dimensions, a portfolio of the form $2 |z_1\rangle \langle \bar z_1 | + |z_2\rangle \langle \bar z_2 |$
consists of a long position of three-halves of one unit of the risk free asset, a long position of one-half of a unit in the A-D security $ \hat \pi_{1}$ and a short position of one-half of a unit in the A-D security $ \hat \pi_{2}$, since we have
$2  \hat \pi_{1} + \hat \pi_{2} = \frac{3}{2} ( \hat \pi_{1} +  \hat \pi_{2} ) + \half ( \hat \pi_{1}   - \hat \pi_{2}) $. In this way, we can  isolate the risk-free part of a portfolio. 
This first notion of a portfolio corresponds rather closely to the notion of a portfolio in a one-period market that arises in classical finance theory \cite{arrow debreu, bingham kiesel, duffie, dothan, etheridge, follmer schied} and can be pursued further in that spirit. The point is that once the measurement basis for the underlying experiment has been fixed, the various associated operators arising for positions with different portfolio weightings  commute.

As we pointed out in Section \ref{sec:Financial observables}, however, it does not make sense to form a portfolio of several contracts each based on the same quantum system but with different measurement frames, since such measurements will in general be incompatible and cannot be simultaneously realized.  
In our approach to the problem, we consider portfolios of assets for which the payouts are based on separate measurements being made on two or more distinct quantum systems. Imagine, for example, a financial institution where in one room an experiment is carried out on Quantum System I, with certain results obtained, and another experiment is carried out in another room on Quantum System II, with certain results obtained. In each case, there are contracts leading to payouts depending on the results obtained. 

Since the measurements do not interfere with one another (after all, they are carried out in different rooms) they can be carried out simultaneously, each delivering a certain number of units of account, so it makes sense to speak of holding a portfolio in the two assets, for which the payout is simply the totality of the payouts of the constituents of the portfolio, with appropriate weightings. 
Let us see how we model such a situation. To simplify the discussion, we stick with the case where there are two quantum systems involved, with measurements made on each of them. 

The setup can then be easily generalized to the case where there are $N$ such systems. The key idea is that to model a portfolio of two such contracts, we need to consider the tensor product of the Hilbert spaces of the individual systems. In fact, the two Hilbert spaces might even be of different dimensions. 

The usual Dirac notation does not hold up so well in such a setting, so we use an {\em index notation} instead, which works quite smoothly \cite{geroch, BH2001}. Thus, let $\mathcal H_1$ be a Hilbert space of dimension $n$ and let $\mathcal H_2$ be a Hilbert space of dimension $n'$, where $n$ and $n'$ are not necessarily the same. We write $\xi^a$ and $\xi^{a'}$ for typical elements of $\mathcal H_1$ and $\mathcal H_2$ respectively, 
where $a = 1, 2, \, .\, .\, .\, n$ and $a' = 1, 2, \, .\, .\, .\, n'$. Thus indices without dashes refer to the first Hilbert space and indices with dashes refer to the second Hilbert space. We write $\eta_a$ and $\eta_{a'}$ for typical elements of the corresponding dual spaces $\mathcal H_1^*$ and $\mathcal H_2^*$. The complex conjugates of  $\xi^a$ and $\xi^{a'}$ are denoted $\bar \xi_a$ and  $\bar \xi_{a'}$ respectively. Then for the inner product between $\xi^a$ and $\eta_a$ we write $\xi^a \,\eta_a$ and for the inner product between $\xi^{a'}$ and $\eta_{a'}$ we write $\xi^{a'} \,\eta_{a'}$, with the usual summation convention.

We are interested in the tensor product Hilbert space 
$\mathcal H_{12} = \mathcal H_1 \otimes \mathcal H_2$, and we write $\xi^{a a'} \in \mathcal H_{12}$ for a typical element of this space. Then we write $\eta_{aa'}$ for a typical element of $\mathcal H_{12}^*$ and $\bar \xi_{aa'}$ for the complex conjugate of 
$\xi^{a a'}$\!, and for the inner product of $\xi^{a a'}$ and $\eta_{aa'}$ we write $\xi^{a a'}\,\eta_{aa'}$. 
The state of a two-particle system takes the form of a density matrix $p^{aa'}_{bb'}$. Thus we require that  it should be Hermitian, of unit trace, and positive, so 
\begin{eqnarray} 
p^{aa'}_{bb'} =  \bar p^{aa'}_{bb'}, \quad   p^{cc'}_{cc'} = 1, \quad p^{aa'}_{bb'} \alpha^b \bar \alpha_a \beta^{b'} \bar \beta_{a'} \geq 0
\end{eqnarray}
for all $\alpha^a, \beta^{a'}$\!. A two-particle density matrix is {\em pure} if $p^{aa'}_{bb'} = \xi^{a a'} \, \bar \xi_{bb'}$ for some state vector  $\xi^{a a'}$. We say that the particles are {\em independent} if  
\begin{eqnarray} \label{independence}
p^{aa'}_{bb'} = p^{a}_{b} \,p^{a'}_{b'} 
\end{eqnarray}
for some pair of one-particle states $p^{a}_{b}$ and $p^{a'}_{b'}$. The state is said to be {\em separable} if it can be written in the form 
\begin{eqnarray} 
p^{aa'}_{bb'} = \sum_{r = 1}^k p^{a}_{b}(r) \,p^{a'}_{b'}(r),
\end{eqnarray}
for some collection of $2k$ one-particle states $ \{p^{a}_{b}(r)\}_{r = 1, 2, \, .\, .\, .\, k}$ and $\{p^{a'}_{b'}(r)\}_{r = 1, 2, \, .\, .\, .\, k}$. But if the two-particle state is not separable then we say that the particles are {\em entangled}. 

Now we are in a position to discuss the idea of measurements on a two-particle system and the contracts one can associate with such measurements. A generic contract based on the outcome of a measurement made on a two-particle system is described by a Hermitian operator $X^{aa'}_{bb'}$\!. We are interested in the case when the measurement splits into a measurement on System I and a measurement on System II and one adds the results to give the payout of the contract. Such a contract takes the form
\begin{eqnarray} \label{two-particle payout}
X^{aa'}_{bb'} =  U^{a}_{b} \delta^{a'}_{b'} +  \delta^{a}_{b} V^{a'}_{b'}\!, 
\end{eqnarray}
where $\delta^{a}_{b}$ and $\delta^{a'}_{b'}$ denote the identity operators on $\mathcal H_1$ and $\mathcal H_2$ respectively. The eigenstates of such an operator are of the form 
\begin{eqnarray} 
p^{aa'}_{bb'} =  \alpha^a \bar \alpha_b \, \beta^{a'} \bar \beta_{b'},
\end{eqnarray}
where $\alpha^a$ is an eigenvector of $U^{a}_{b}$ and $\beta^{a'}$ is an eigenvector of $V^{a'}_{b'}$. Thus 
$U^{a}_{b} \alpha^b = u \alpha^a$ and  $V^{a'}_{b'} \beta^{b'} = v \beta^{a'}$ for $u, v \in \mathbb R^+$ and the sum $u + v$  gives the overall payout of the contract. Such a contract represents a portfolio consisting of one unit of
a contract based on System I and one unit of a contract based on System II. More generally, for a portfolio consisting of $\theta_1$ units of the first contract and $\theta_2$ units of the second contract we have
\begin{eqnarray} 
X^{aa'}_{bb'}(\theta_1, \theta_2) =  \theta_1 \,U^{a}_{b} \delta^{a'}_{b'} +  \theta_2\,\delta^{a}_{b} V^{a'}_{b'} 
\end{eqnarray}
and the payout will be of the form $ \theta_1u + \theta_2v$. The setup for a portfolio of arbitrary size can be constructed analogously. In particular, one can check that the expected payout of a portfolio is equal to the sum of the expectations of the constituents. This is because whenever the density matrix of the two-particle state hits one of the identity operators in the portfolio operator, all but one of the systems gets traced out and one is left with the trace of the product of a single particle density operator and the observable associated with that system. For example, in the case of a two-particle system one finds that
\begin{eqnarray} 
p_{aa'}^{bb'} X^{aa'}_{bb'}(\theta_1, \theta_2) &=& p_{aa'}^{bb'} ( \theta_1 \,U^{a}_{b} \delta^{a'}_{b'} +  \theta_2\,\delta^{a}_{b} V^{a'}_{b'} )
\nonumber \\
&=& \theta_1 \,p_{aa'}^{bb'}  \,U^{a}_{b} \delta^{a'}_{b'} +  \theta_2\,p_{aa'}^{bb'}  \delta^{a}_{b} V^{a'}_{b'} 
\nonumber \\
&=& \theta_1 \,p_{a}^{b} \,U^{a}_{b} +  \theta_2\,p_{a'}^{b'} V^{a'}_{b'},
\end{eqnarray}
where $p_{a}^{b} = p_{ac'}^{bc'}$ and $p_{a'}^{b'} = p_{ca'}^{cb'}$. Likewise one can check that the price of a portfolio is equal to the weighted sum of the prices of its constituents. The point is that the two-particle system is itself a quantum system with a financial observable based on it, of the form \eqref{two-particle payout}, so by  Proposition \ref{prop:1} there exists a pricing operator $q_{aa'}^{bb'}$ such that 
\begin{eqnarray} 
q_{aa'}^{bb'} X^{aa'}_{bb'}(\theta_1, \theta_2) =  \theta_1 \,q_{a}^{b} \,U^{a}_{b} +  \theta_2\,q_{a'}^{b'} V^{a'}_{b'},
\end{eqnarray}
where the traced-out operators $q_{a}^{b} = q_{ac'}^{bc'}$ and $q_{a'}^{b'} = q_{ca'}^{cb'}$ are the pricing operators associated with the respective individual systems.

There is one further aspect of the portfolio problem that can be analyzed and this concerns the matter of correlations. If the state of the two-particle system is of the form \eqref{independence}, so the two particles are independent, then the outcomes of the experiments on the two systems will be uncorrelated. 
But if the systems are entangled, then the correlation will in general be non-vanishing, leading to relations such as 
\begin{eqnarray} 
p_{aa'}^{bb'} \big [U^{a}_{b} - \delta^{a}_{b} \,p_{c}^{d} \,U^{c}_{d}\big] \big [V^{a'}_{b'} - \delta^{a'}_{b'} \,p_{c'}^{d'} \,V^{c'}_{d'}\big] \neq 0.
\end{eqnarray}
The point about entanglement is that even if the two systems are in separate rooms (or even different cities) the outcomes may be correlated, owing to the original construction of the state of the two-particle system to which they belong. The same is true of the prices: if  $q^{aa'}_{bb'}$ is entangled, then there will be correlations in the prices, as shown in relations such as
\begin{eqnarray} 
q_{aa'}^{bb'} \big [U^{a}_{b} - \delta^{a}_{b} \,q_{c}^{d} \,U^{c}_{d}\big] \big [V^{a'}_{b'} - \delta^{a'}_{b'} \,q_{c'}^{d'} \,V^{c'}_{d'}\big] \neq 0.
\end{eqnarray}

Thus, in the general situation we see that when there is a market based on contracts associated with measurements being made on a number of different quantum systems, there will be correlations between outcomes of measurements and correlations between prices, where the former are determined by the structure of physical density operator for the market as a whole and the latter by the structure of the pricing operator for the market as a whole. 

\section {Conclusion}
\label{sec:Conclusion} 

\noindent The physical density operator is objective in nature, the only limitations in its determination being in
the usual practicalities of the laboratory settings where the states are manufactured. 
The pricing operator, on the other hand, if classical finance theory is any guide in the matter \cite{arrow debreu, bingham kiesel, dothan, duffie, etheridge, follmer schied}, will be determined by the collective appetite for risk and reward among market participants. Hence, as in all markets, prices will be subject to fluctuation and change over time and may even be amenable to a Bayesian treatment. In the one-period setting that we have developed here, all we can say {\em a priori} of a definite nature about the pricing operator is that it exists and that the physical density  operator and the pricing operator are equivalent, as we have seen in Proposition \ref{prop:1}. 

In the one-period version of the theory, one can be somewhat agnostic on the matter of dynamics. This is because $\hat p$, $\hat q$ and $P_{0T}$ are specified at time $0$ and no further data are needed apart from the observable $\hat X_T$ being measured at time $T$. From a dynamical perspective it is convenient to work in the Heisenberg representation. Then $\hat p$ and $\hat q$ are fixed and $\{ \hat X_t\}_{t\geq0}$ is dynamical, given by 
\begin{eqnarray} \label{Heisenberg dynamics}
\hat X_T = \re^{{-\rm i} \hat H T} \, \hat X_0\, \re^{{\rm i} \hat H T},
\end{eqnarray}
where $\hat X_0$ denotes the initial value of the observable being measured and $\hat H$ is the Hamiltonian of the underlying physical system. Since $T$ is fixed, it suffices to specify $\hat X_T$, and we let $\hat X_0$ and $\hat H$ drop out of the picture. 

The Heisenberg representation is also convenient when interventions are taken into account in a multi-period model. Suppose, for example, we consider a two-period model involving a pair of systems defined on the product of two Hilbert spaces. We write $0 < t < u$ and let $X^a_b (t)$ and $Y^{a'}_{b'} (u)$ be a pair of observables, one for a measurement acting on the first particle at time $t$ and another for a measurement acting on the second particle at time $u$. If we assume that the two particles are non-interacting, then the two observables evolve independently, each according a law of the form \eqref{Heisenberg dynamics}, with distinct Hamiltonians. The physical state of the two-particle system can be represented in line with the scheme outlined in the previous section by a tensor of the form $p_{aa'}^{bb'}(0)$ and for the pricing state we write $q_{aa'}^{bb'}(0)$. Note that although the two particles are non-interacting, we allow for the possibility that they may have been prepared in an entangled state, so the two density matrices need not be separable. 

Then for the valuation of the contract defined by the measurement of $X^a_b (t)$ alone, one can work with the reduced density matrices defined by $p_{a}^{b}(0) = p_{ac'}^{bc'}(0)$ and $q_{a}^{b}(0) = q_{ac'}^{bc'}(0)$. But for a payout involving both a measurement of $X^a_b (t)$ and a measurement of $Y^{a'}_{b'} (u)$, matters are a little more complicated. This is because once the result of the first measurement is known, the market may change its assessment of the pricing state, in line with classical idea that the pricing kernel is an adapted process, so that market participants will adjust their attitudes towards risk following a movement in the market. We refer to such state changes in the Heisenberg representation as ``interventions." Now, the change in the physical state is relatively straightforward: this is the usual L\"uders state-reduction rule \cite{luders}, which depends on the outcome of the measurement of $X^{a}_{b} (t)$. The transformation is thus given by
\begin{eqnarray} 
p_{aa'}^{bb'}(0) \to p_{aa'}^{bb'}(t) = L_{a}^{c}\, L_{d}^{b} \, p_{ca'}^{db'}(0) /  L_{e}^{c}\, L_{d}^{e} \, p_{cf'}^{df'}(0),
\end{eqnarray}
where $L_{a}^{b}$ denotes the projection operator onto the Hilbert subspace defined by the random outcome of the measurement of $X^{a}_{b} (t)$. 

But the pricing state need not follow the L\"uders rule: it is constrained only by the requirement that the new pricing state arising after the measurement should be equivalent to the new physical state. Thus, in general, we have a transformation of the form 
\begin{eqnarray} 
q_{aa'}^{bb'}(0) \to q_{aa'}^{bb'}(t),
\end{eqnarray}
where $q_{aa'}^{bb'}(t)$ and $p_{aa'}^{bb'}(t)$ share the same null space.  One way of achieving this is by taking any alternative state $r_{aa'}^{bb'}(0)$ which is equivalent to $p_{aa'}^{bb'}(0)$ and then passing it through the
L\"uders projection sieve  on the first Hilbert space to give 
\begin{eqnarray} 
q_{aa'}^{bb'}(t) = L_{a}^{c}\, L_{d}^{b} \, r_{ca'}^{db'}(0) /  L_{e}^{c}\, L_{d}^{e} \, r_{cf'}^{df'}(0).
\end{eqnarray}
 Then for each possible result of the first measurement we obtain a new pricing state, which can be used to form a time-$t$ conditional valuation of the payout triggered by the later time-$u$ measurement. This can be compared with the time-$0$ valuation of the second payout, which is obtained by using the original 
 pricing state but tracing out the first Hilbert space. 
 
 Thus, one can think of the structured product under consideration as a contract with two cash flows, one at $t$ and one at $u$. The value of the contract at time 0 is thus
\begin{eqnarray}\label{pricing formula}
S_0 = P_{0t} \,  q_{aa'}^{bb'}(0) \, X^a_b (t) \,\delta^{a'}_{b'} + P_{0u} \, q_{aa'}^{bb'}(0) \,\delta^a_b \,Y^{a'}_{b'} (u).
\end{eqnarray}
Then at time $t$ the contract delivers its first cash flow and goes ex-dividend; and its new value, conditional on the outcome of the first measurement, is
\begin{eqnarray}\label{pricing formula}
S_t =  P_{tu} \, q_{aa'}^{bb'}(t) \, \delta^a_b \, Y^{a'}_{b'} (u),
\end{eqnarray}
where  $P_{st} = P_{0t} / P_{0s}$ is the usual forward discount factor. Note that $q_{aa'}^{bb'}(t)$ depends on the random outcome of the first measurement. Finally, the second cash flow kicks in at time $u$ and the asset goes ex-dividend once more, so we have $S_u = 0$. In this way we obtain stochastic processes for the value of the asset and its dividend flow.  The scheme can easily be generalized to a market of any number of periods, each involving a new measurement.
 
That the non-Kolmogorovian character of quantum probability may have implications for the development of quantum technologies is widely appreciated -- see \cite{holik} and references cited therein. And indeed, if quantum computers eventually replace the classical computers currently used for algorithmic trading by financial institutions, as they no doubt will, then the role of valuations of the type we have considered here may be important in that context. 
There is also a widely held view that quantum probability may play a part in cognitive science and hence behavioural finance as well -- see \cite{brody 2023, busemeyer bruza 2012, khrennikov 2010, haven khrennikov 2016, pothosbusemeyer2022, pothos busemeyer 2013, yukalov sornette 2017} and references cited therein. The suggestion is that the brain uses quantum probability in a crucial way in its decision-making apparatus. In that respect, quantum cognition and quantum psychology can be viewed as a promising basis through which asset prices might be subject to quantum laws.
It would be outside of the scope of the present discussion to look at such proposals in detail here, but  if judgements and decisions are made on the basis of quantum probability, then in some situations these assessments will involve {\em valuations}, rather than probability estimates, and it would be the pricing operator, rather than the physical density operator, that would come into play in these valuations. In such cases, external intervention in the form of Bayesian updating could be modelled, e.g., as in \cite{brody 2023}. This is consistent with the point we made earlier about the pricing operator being specific to the risk and reward profiles of market operatives and in a state of flux as new information arrives. These and other further developments of the theory we hope to explore elsewhere. 



\newpage

\noindent {\bf APPENDIX: Quantum Theory, Finance Theory, and Quantum Computing}
\vspace{0.5cm}

\noindent In Section \ref{sec:Introduction}, we mentioned two rather distinct categories of investigation being pursued under the general heading of quantum finance. Category I consists of theories that explore the idea that asset prices are subject to quantum laws. Category II involves the application of mathematical techniques that have arisen in the context of quantum theory to problems in computational finance. The work of Segal \& Segal \cite{segal1998} is an good example in the first category. Further examples can be found in \cite{baaquie2013, B D'Hooghe, chen2004, Focardi2020, haven2005, haven khrennikov 2016, khrennikov 2010, McCloud2018, bao rebentrost}, to mention but a few. 
The problem with this line of thinking is that there is no direct evidence to suggest that asset prices are in any general sense ``quantum-like" in nature, so work along these lines is speculative. Even if one were to admit the idea that an asset price (say, the price of a barrel of oil) is akin to a physical variable, such as position or energy, that can be quantized, it is not at all evident what form the associated complementary variable would take. And where would Planck's constant enter the discussion?  

One way around these issues is the suggestion that the brain uses quantum probability in its decision-making apparatus, even if its physical structure is essentially classical. In that respect, so-called quantum cognition and quantum psychology can be viewed as a promising basis through which asset prices might be subject to quantum laws~\cite{busemeyer bruza 2012, khrennikov 2010, haven khrennikov 2016, pothosbusemeyer2022, pothos busemeyer 2013, yukalov sornette 2017}. 

In the second category, involving applications of quantum methods to the solution of ``classical" problems in finance, including the use of quantum computation for this purpose, there is a very considerable literature, and one can mention \cite{assouel2022, baaquie2013, baaquie2018, fedorov2022, herman, Jacquier2022, Rebentrost2018, stamatopoulos2022} as representative of the multiplicity of ideas being pursued. 

Since our work overlaps, to some extent, with that of Bao \& Rebentrost \cite{bao rebentrost} (B-R), we comment on points where we agree and where we differ. B-R introduce the idea of a market described by a quantum density operator rather than by classical probabilities, then introduce a class of so-called quantum assets. A quantum asset is defined to be a positive semidefinite Hermitian matrix. The assets are given a ``financial interpretation" by B-R, who say that,\,``Each eigenstate can be considered a natural event for the quantum asset.  Each eigenvalue is the outcome or payoff of this asset when the corresponding event happens." 
We are in agreement with the substance of the idea here, though in our approach there is no abstraction in the definition: a quantum asset is a financial contract with a well-defined structure involving a payoff contingent on an experiment.  In contrast, of their definition B-R say,\,``This definition leaves questions about the existence/validity of such assets and their intrinsic value for future work."  B-R give a ``toy example" involving a market maker who has access to a quantum computer. The idea is that initially the state is known to the market and the current bid and offer prices for a certain asset. Then a unitary transformation, also known to the market, is applied to the state and a measurement is made, the random outcome of which determines the new bid and offer prices made by the market maker.  Investors at the initial prices then may or may not make a profit by trading again at the new prices. While such a setup may be feasible, it cannot be said that the payoff of the asset is determined by a quantum computation, since the algorithms used by quantum computers are typically designed to give a definite or near-definite result, not a probabilistic result -- so the introduction of a quantum computer in B-R's toy example is spurious. In any case, no logical basis is provided on which a market maker would make a random price on the back of a quantum measurement. Part of the problem here is that B-R fail to make a clear distinction between the price and the payout of an asset, and this leads to the confused idea that a market maker makes a random price based on the outcome of a measurement. 

We also differ from B-R in our treatment of the no-arbitrage condition. B-R introduce their idea of a portfolio of quantum assets as a linear combination of the matrices associated with the various assets, each matrix being weighted by the number of units held in that asset. The ``expected value" of the portfolio is then given (according to B-R) by the trace of the product of the market density operator and the weighted sum of the matrices. The problem is that the weighted sum of the assets is indeed an asset itself, but its payoffs are not given by weighted sums of the payoffs of the individual assets. In plain language, if one forms the sum of two matrices, then the eigenvectors and eigenvalues of the sum are related to those of the individual matrices only rather indirectly. In this respect, our approach differs completely from that of B-R, who attempt to set up a portfolio theory, and then a theory of arbitrage, based on the idea that all of the assets are associated with the same Hilbert space. In fact, B-R give no clear statement of what is meant in real terms by a ``quantum asset" in their theory or by a portfolio of such assets.
We conclude that B-R's attempt to propose an analogue of the fundamental theorem of asset pricing is ill-posed.  Moreover, B-R simply assume the existence of a pricing state (in the form of a density matrix) rather than deducing its existence.  

In our theory of quantum finance, the assumption of no arbitrage is that of Axiom 1; the relation to classical finance is embodied in Axiom 2; and the risk-free rate is specified in Axiom 3. The physical density operator $\hat p$ arises in the specification of the experiment that underlies the contract. The density operator $\hat q$ is shown to exist in Proposition 1 and its equivalence to $\hat p$ follows as a consequence of Axiom 1.

\begin{acknowledgments}
\noindent
The authors wish to thank D C Brody and B K Meister for helpful comments.  
\end{acknowledgments}

\vspace{0.50cm} 

\noindent {\bf References}



\end{document}